\documentclass[11pt]{article}
\usepackage{moriond,epsfig}

\def\Mkk{M_{\rm KK}}
\def\be{\begin{equation}}
\def\ee{\end{equation}}
\def\bea{\begin{eqnarray}}
\def\eea{\end{eqnarray}}

\newcommand{\ord}{{\cal O}}
\begin{document}
\vspace*{4cm}
\title{HIGGS PHYSICS IN WARPED EXTRA DIMENSIONS}

\author{ FLORIAN GOERTZ }

\address{Institut f\"ur Physik (THEP), Johannes Gutenberg-Universit\"at\\
 D-55099 Mainz, Germany}

\maketitle\abstracts{
In this talk, I present results for the most important Higgs-boson production cross sections at the LHC and the Tevatron as well as the 
branching fractions of the relevant decay channels in the custodial Randall-Sundrum model. The results are based on a complete
one-loop calculation, taking into account all possible Kaluza-Klein particles in the loop. Due to the strong infrared 
localization of the top quark and the Kaluza-Klein excitations, the SM predictions receive sizable corrections in the model at hand. 
This could effect Higgs searches significantly.}

\section{Introduction}

The Higgs boson represents the last missing ingredient of the Standard Model (SM) of Particle Physics. It offers the possibility 
to give masses to the weak gauge bosons and chiral fermions without breaking gauge invariance, which is important for a proper 
high energy behavior of the model. Electroweak precision measurements suggest that the SM Higgs boson is light, 
$m_h < 185$\,GeV at 95\%\,C.L.,\cite{:2010vi} including the direct Limit $m_h>114$\,GeV from LEP2. Furthermore, theoretical 
arguments like unitarity, vacuum stability and triviality constrain the allowed range for the Higgs mass. In summary, 
we expect the SM Higgs boson to have a mass well below a TeV and to exhibit tree-level couplings to particles proportional 
to their mass. Imagine we do not discover the Higgs at the Large Hadron Collider (LHC) in the first years of running. 
Does this already mean that we have to abandon the corresponding mechanism of electroweak symmetry breaking? The answer is 
certainly no. Beyond the SM physics could feature a standard Higgs mechanism that could be much harder 
to detect, even for a Higgs mass easily accessible at the LHC. It is important to study Higgs physics in various models
to be prepared for different possible scenarios. In this talk, I present results for Higgs production and decay 
within the custodial Randall-Sundrum (RS) model with gauge and fermion fields in the (5D) bulk and an infrared-brane Higgs sector. 
Here one expects big effects, due to the localization of the fields. 

\section{Aspects of the Randall-Sundrum Model}
The RS model \cite{Randall:1999ee} provides an elegant possibility to address the large hierarchy between the electroweak 
scale $M_{\rm EW}$ and the Planck scale $M_{\rm PL}$ by means of a non-trivial geometry in a 5D Anti-de Sitter space. The 
fifth dimension is compactified on an $S^1/Z_2$ orbifold. The RS metric
\be
ds^2=e^{-2 kr|\phi|}\eta_{\mu\nu}\,dx^\mu dx^\nu-r^2d\phi^2\,,
\ee
with $\eta_{\mu\nu}=\,$diag(1,-1,-1,-1) is such that length scales within the usual 4D space-time are rescaled by an 
exponential warp factor, depending on the position $\phi \in [-\pi,\pi]$ in the extra dimension. The curvature $k$ and inverse 
radius $r^{-1}$ of this dimension are of $\ord\left(M_{\rm PL}\right)$. The $Z_2$ fixed points at $\phi=0,\pi$ correspond 
to boundaries: the ultraviolet (UV) and the infrared (IR) 3-branes. The model solves the gauge hierarchy 
problem by suppressing mass scales on the IR-brane.~One~achieves
\be
M_{\rm IR}\equiv e^{-L} M_{\rm Pl} \approx M_{\rm EW}\,
\ee
for $L\,\equiv kr\pi\approx 36$. The strong hierarchy between $M_{\rm PL}$ and $M_{\rm EW}$ is thus understood by gravitational 
red-shifting, if the Higgs field is localized on or close to the IR brane. The 5D gauge and fermion fields are decomposed into 
infinite towers of (massive) 4D fields, featuring profiles depending on $\phi$, via a Kaluza-Klein (KK) decomposition. 
The massless zero modes can become massive via couplings to the Higgs sector and, given an appropriate setup of the model,
they can be interpreted as the SM fields we observe in nature. The compactification of the fifth dimension leads to masses 
for the tower of KK excitations, which are set by the KK scale $\Mkk\equiv k \epsilon \sim \ord({\rm TeV})$. 
The warping of the fundamental Planck scale down to $M_{\rm IR} \sim \ord({\rm TeV})$ on the IR brane results in a 
cutoff for the RS model at several TeV for amplitudes calculated on that brane. At this scale the model is assumed 
to be UV completed by a theory of quantum gravity. This is important for Higgs physics, because it means that just the 
exchange of the first KK excitations should be taken into account for the corresponding observables, while the effect 
of the higher modes is to be cut off. An attractive feature of RS models is the possibility to address the quark-mass 
hierarchies and the structure of the Cabibbo-Kobayashi-Maskawa (CKM) matrix by localizing the fermion zero modes differently in 
the extra dimension, without any hierarchies in the input parameters.\cite{Huber:2000ie} 
This anarchic approach to flavor improves the predictivity of the model, since the localizations
of the quarks are now fixed to some extend by their masses and the CKM parameters. The top quark, being the heaviest 
quark of the SM, has to reside close to the IR brane, where also KK excitations tend to live. Due to the 
large overlap with these excitations, one expects the most interesting signatures of RS models in top and Higgs physics. 
A direct consequence of the different localizations of fermions are flavor changing neutral currents (FCNCs). The 
non-universal couplings to massive gauge bosons lead to offdiagonal transitions, after going to the mass basis. Furthermore, 
the KK masses (which are due to compactification) lead to a misalignment between the mass and the Yukawa matrix
which results in modified Higgs couplings for RS models including FCNCs. Our SM assumption for Higgs searches,
a coupling given by the mass of the corresponding particle, is spoiled in this model. The most optimistic 
RS predictions for ${\cal B} (t \to c Z)$ and ${\cal B} (t \to c h) $ are both around $10^{-5}$.\cite{Casagrande:2008hr,Casagrande:2010si}
Let me finally mention that in the {\it minimal} RS model a leading order analysis of the electroweak $S$ and $T$ parameters 
favors a heavy Higgs boson $m_h \sim 1$\, TeV (which is not true for the custodial version) and that the theoretical 
Higgs-mass bounds can be altered. For more details and further references see \cite{Casagrande:2008hr}.  
The custodial RS model that provides the framework for the following analysis of Higgs physics, which is based on \cite{Casagrande:2010si}, 
features a protection for the $T$ parameter \cite{Agashe:2003zs} as well as for $Z b_L \bar b_L$ couplings.\cite{Agashe:2006at}

\section{Higgs Production}

\begin{figure}[!t]
\begin{center} 
\hspace{-2mm}
\mbox{\includegraphics[height=2.3in]{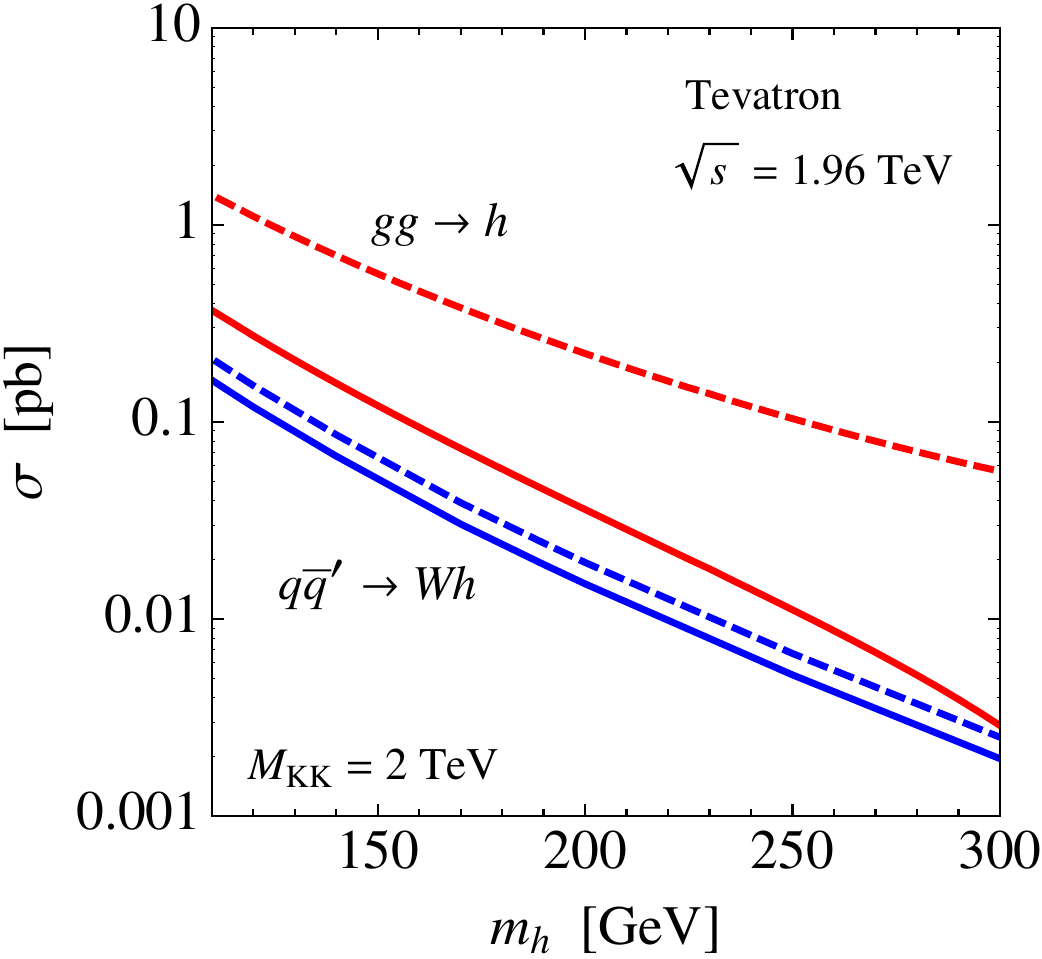}} 
\hspace{2mm}
\mbox{\includegraphics[height=2.3in]{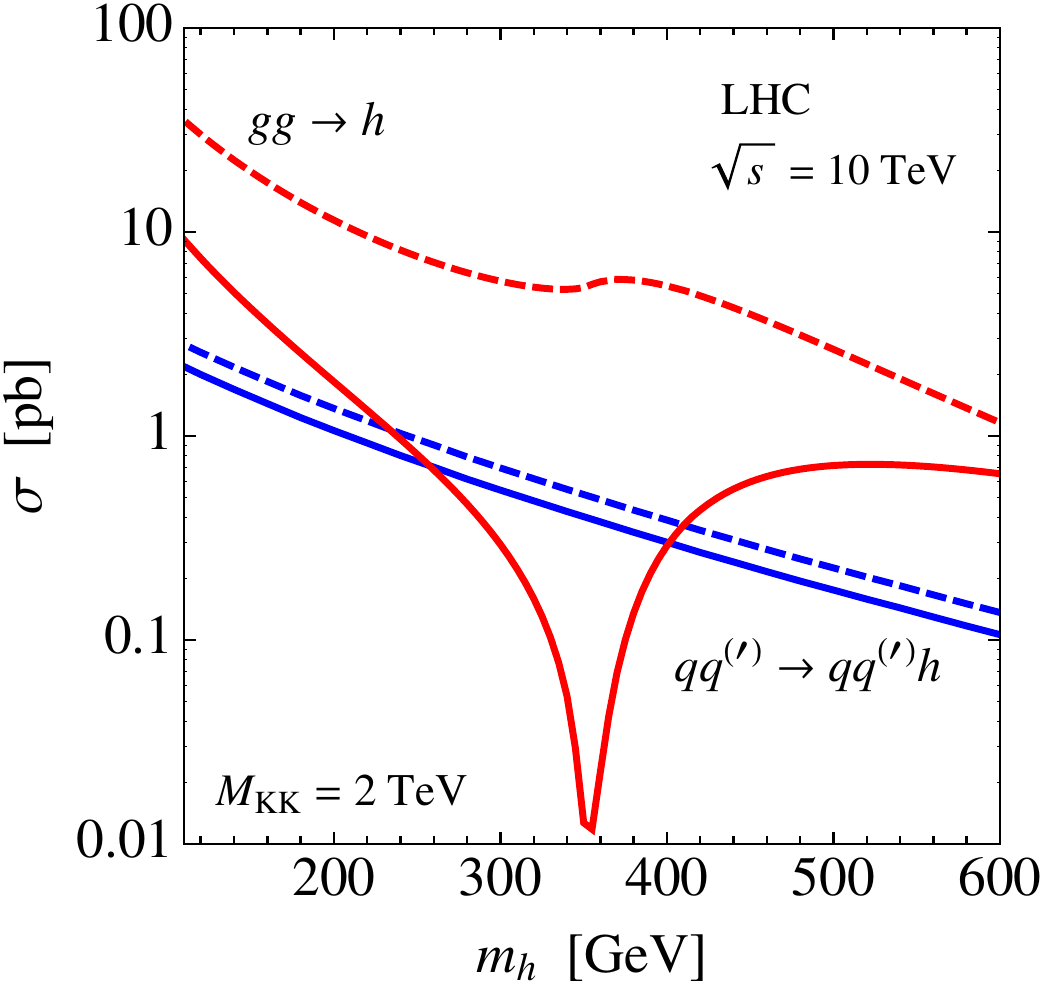}}
\caption{\label{fig:prodplots} Main Higgs-boson production cross sections 
	at the Tevatron (left) and the LHC (right). The dashed lines represent the SM predictions, 
	while the solid lines correspond to the custodial~RS~model. See text for details.}
\end{center}
\end{figure}

The main production mechanism of the Higgs boson at hadron colliders 
is gluon-gluon fusion. In the SM, this process
is dominated by a top-quark triangle loop. Within RS models, one has 
to consider additionally the KK tower of the top quark as well as of all other 
flavors present in the theory, which all contribute at the same order. 
The corresponding Feynman diagrams are given on the very left in the top row 
and bottom row in Figure~\ref{fig:hXX}. In order to obtain the $gg \to h$ 
production cross section in the custodial RS model, the SM prediction is 
rescaled according to
\be \label{eq:rescale1}
\sigma (gg \to h)_{\rm RS} = \left | \kappa_{g} \right |^2 \,
\sigma (gg \to h)_{\rm SM} \,,\qquad
\ee
where
\be \label{eq:kappagg} 
  \kappa_{g} = \frac{{\displaystyle
      \sum}_{i = t, b} \, \kappa_i \hspace{0.25mm} A_{q}^h
    (\tau_i)\, + \hspace{0.25mm}{\displaystyle \sum}_{j = u,d,\lambda}  
    \, \nu_j }{ {\displaystyle \sum}_{i = t,
      b} \; A_{q}^h (\tau_i)} \,,
\ee
with $\tau_i \equiv 4 \hspace{0,25mm} m_i^2/m_h^2$. The first sum in the numerator contains top and bottom quark zero modes 
running in the loop with couplings (normalized to the SM values) given by
\be
\kappa_t={\rm Re}[(g_h^u)_{33}]/\left(\frac{m_t}{v}\right)\,,\quad \kappa_b={\rm Re}[(g_h^d)_{33}]/\left(\frac{m_b}{v}\right).
\ee
Here, $v\approx 246$\, GeV is the Higgs vacuum expectation value and $m_t$ ($m_b$) is the top (bottom) quark mass. 
The Higgs couplings $(g_h^{u,d})_{33}$ in the custodial RS model as well as the form factor 
$A_{q}^h (\tau_i)$ can be found in \cite{Casagrande:2010si}. It is easy to show that in the RS model $\kappa_{t,b}<1$, 
independent of the input parameters,\cite{Casagrande:2010si} where $\kappa_t$ can become as small as 0.5 for $\Mkk=2$\, 
TeV, which we will always employ in the following analysis. The second sum in (\ref{eq:kappagg}) represents the contribution 
from the virtual exchange of KK excitations. The $\lambda$ quarks, with electromagnetic charge $5/3$, arise in the custodial 
RS model due to the more complicated fermion structure in order to protect the $Z b_L \bar b_L$ vertex. 
Details on the sums over KK excitations are given in \cite{Casagrande:2010si}.
Note that the contributions of the first KK levels (after summing the different same-charge flavors within 
a level) turn out to decrease quadratically. Thus the extrapolation from these levels to the whole tower, 
which actually should be cut off, does not change the results significantly. 
The results for the Higgs-boson production cross sections at Tevatron and the LHC 
for center-of-mass energies $\sqrt{s} = 1.96 \, {\rm TeV}$ and $\sqrt{s} = 10 \, {\rm TeV}$ are shown in
Figure~\ref{fig:prodplots}. The solid red lines correspond to the custodial RS expectations, whereas the SM 
predictions are indicated by dashed lines for comparison. In addition to gluon-gluon fusion, the 
plots show (in blue) the predictions for weak gauge-boson fusion, $q q^{(\prime)} \to qq^{(\prime)} V^{\ast}
V^{\ast} \to qq^{(\prime)} h$ with $V = W,Z$, which is an important channel at the LHC, as well as for associated 
$W$-boson production, $q \bar q^{\hspace{0.25mm} \prime} \to W^\ast \to Wh$, for the Tevatron. 
The results have been obtained by an averaging procedure over 10000 sets of
input parameters, fitting the quark masses, CKM mixing angles and the phase within $1\sigma$.
The plots show clearly that the Higgs production cross sections in gluon-gluon
fusion experience a strong reduction in the custodial RS model. This depletion remarkably survives
even for $\Mkk=5$\, TeV, which corresponds to a first KK-gauge boson mass of around 12 TeV, still reaching 
up to $-40\%$ for both colliders. The bump in the right plot is due to a destructive interference of the zero
mode and KK contributions which becomes most effective for $m_h\approx 2 m_t$.

\section{Higgs Decay}

\begin{figure}[!t]
\hspace{-0.6cm}
\raisebox{2.4cm}{
\begin{tabular}[!t]{l}
\includegraphics[width=6.6cm]{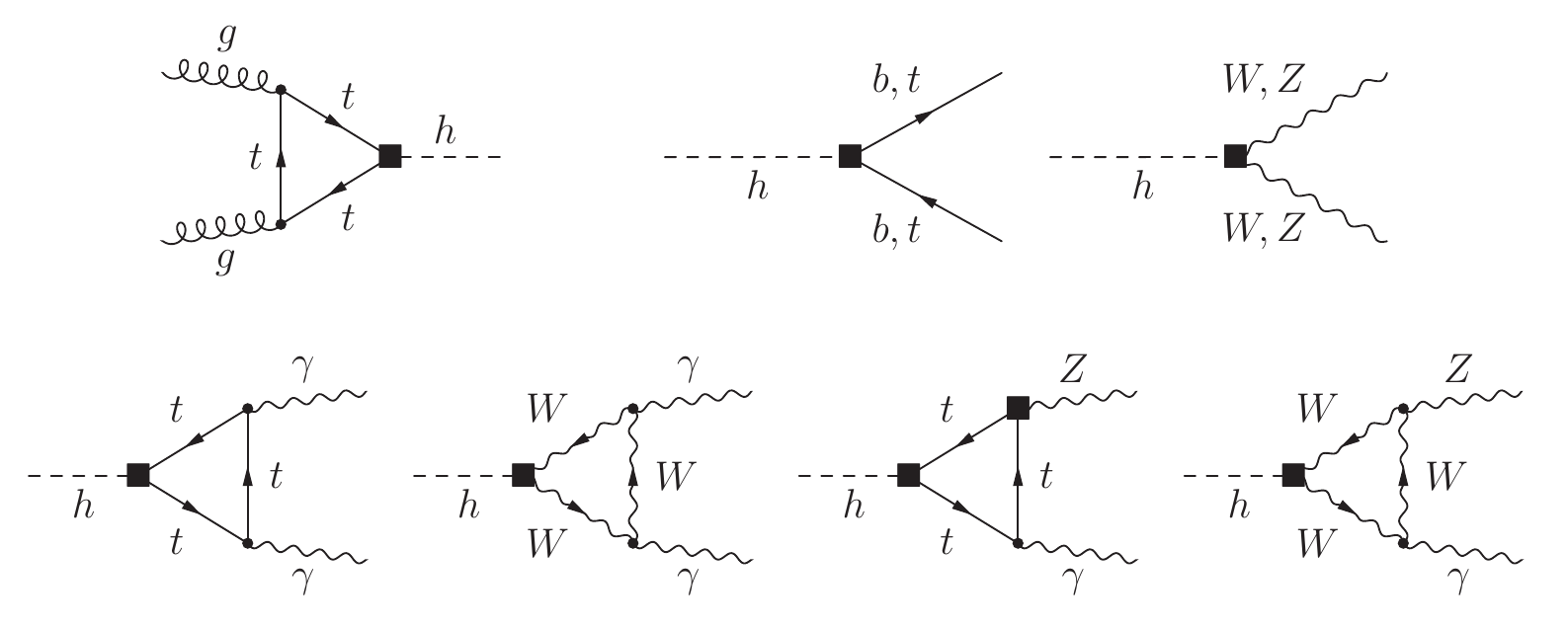}\\
\includegraphics[width=6.1cm]{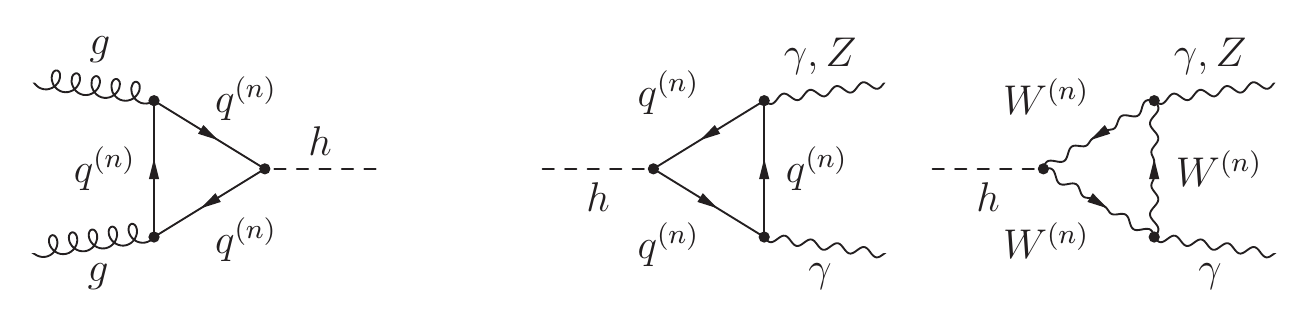}
\end{tabular}}\quad
\includegraphics[width=9.1cm]{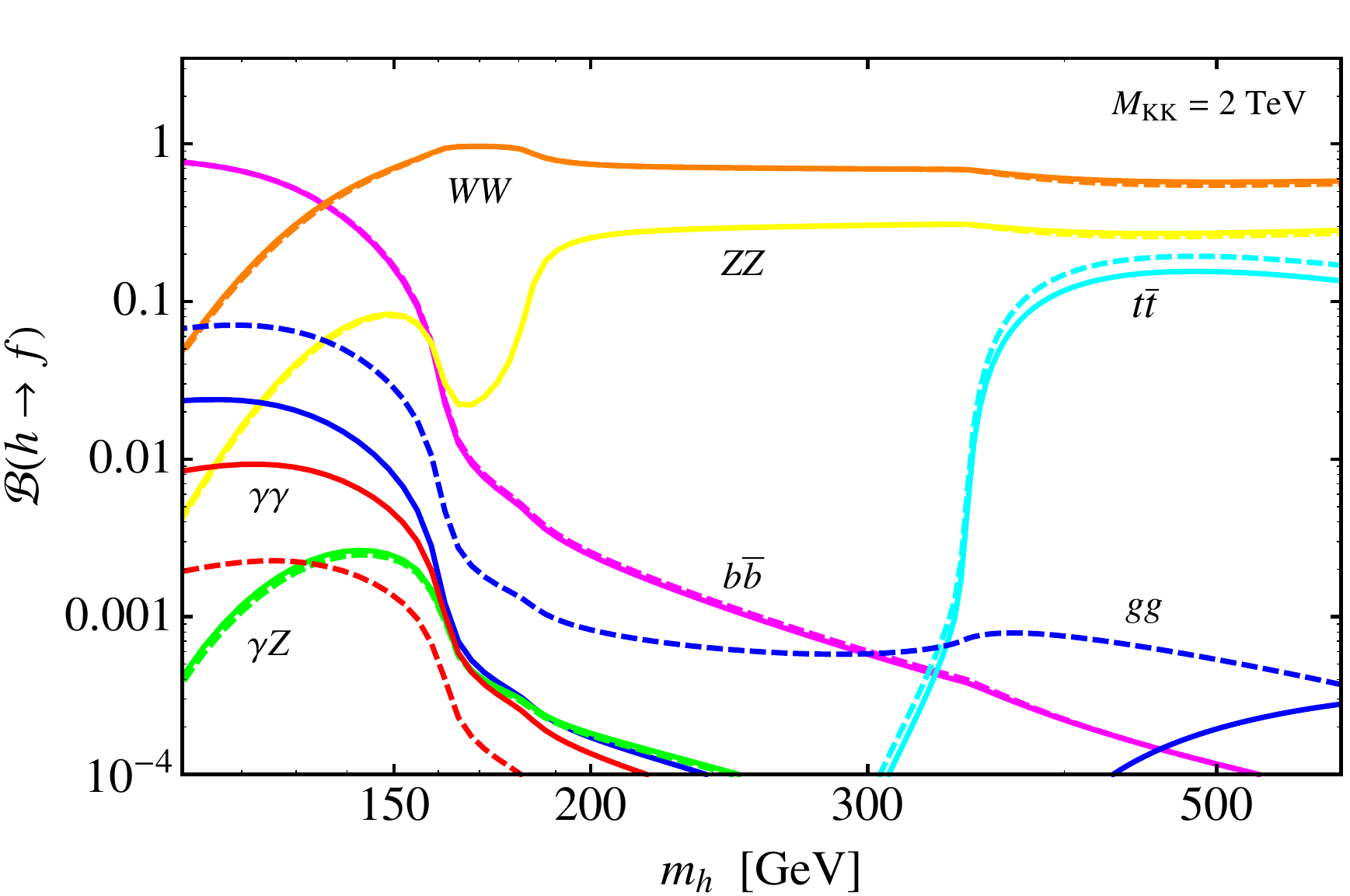}
\caption{\label{fig:hXX}left: Feynman diagrams for Higgs production and decay, right: Branching ratios for $h \to
    f$ as functions of the Higgs-boson mass. The
    solid (dashed) lines indicate the custodial RS (SM) predictions. See text for details.}
\end{figure}
Concerning the decay of the Higgs boson, processes with heavy quarks and gauge bosons in the final state 
can experience significant RS corrections. The corresponding Feynman diagrams are depicted on the left of Figure~\ref{fig:hXX}, where
vertices that receive non-negligible corrections are indicated by black squares. 
The analysis works in a similar way as that for Higgs production.\cite{Casagrande:2010si}
The results are shown on the right of Figure~\ref{fig:hXX}, where the solid (dashed) lines correspond to the RS (SM) predictions. 
All final states that can feature non-negligible RS corrections and
have branching fractions above $10^{-4}$ are considered. While for Higgs masses below the $WW$ threshold the enhanced 
branching fraction into two photons could compensate the lower production cross section in $gg\rightarrow h 
\rightarrow \gamma\gamma$, the discovery potential above this threshold is for all channels significantly 
worse than in the SM. Most important, the golden channel $gg\rightarrow h \rightarrow Z^{(*)} Z^{(*)}\rightarrow l^+l^-l^+l^-$ 
suffers from the strong reduction in the production cross section.
The presented results suggest that a discovery of the Higgs boson, depending on its mass, could become more difficult 
in RS models. Existing SM bounds on the Higgs mass from the Tevatron and LEP are also altered if warped extra dimensions are realized 
in nature. Furthermore, the effects in Higgs physics should be notable at the LHC, even for KK scales which are by far 
not directly accessible. 

\section*{Acknowledgments}
I would like to thank the organizers of the {\it Rencontres de Moriond 2011 EW} for the wonderful atmosphere during the conference
and the financial support. Furthermore, I want to thank Sandro Casagrande, Uli Haisch, Matthias Neubert, and Torsten Pfoh for the 
nice collaboration on the subject and Julia Seng for carefully proofreading the manuscript.

\end{document}